\documentclass[greybox]{svmult}

\usepackage{mathptmx}       
\usepackage{helvet}         
\usepackage{courier}        
\usepackage{type1cm}        
\usepackage{makeidx}         
\usepackage{graphicx}        
\usepackage{multicol}        
\usepackage[bottom]{footmisc}

\usepackage{graphicx}
\usepackage{stfloats}

\begin{document}

\title*{Tracking bitcoin users activity using community detection on a network of weak signals}

\author{Cazabet Remy, Baccour Rym, Latapy Matthieu}

\institute{Cazabet R.\at Univ Lyon, UCBL, CNRS, LIRIS UMR 5205, F-69621, Lyon, France, \email{remy.cazabet@gmail.com}
\and Baccour R., Latapy M., Cazabet R. \at Sorbonne Universit\'es, UPMC Univ Paris 06, CNRS, LIP6 UMR 7606, 75005 Paris France}

\maketitle

\abstract{ Bitcoin is a cryptocurrency attracting a lot of interest both from the general public and researchers. There is an ongoing debate on the question of users' anonymity: while the Bitcoin protocol has been designed to ensure that the activity of individual users could not be tracked, some methods have been proposed to partially bypass this limitation. In this article, we show how the Bitcoin transaction network can be studied using complex networks analysis techniques, and in particular how community detection can be efficiently used to re-identify multiple addresses belonging to a same user.
}

\newtheorem{heur}{Heuristic H}

\section{Introduction}
Bitcoin is a cryptocurrency, a digital currency using a technology called \textit{blockchain} to maintain a public distributed ledger. Bitcoin is decentralized, i.e. there is no central authority. Instead, any peer can consult the distributed ledger at any time to check the amount of b	itcoin controlled by each \textit{address}, which is in fact a public key.

The bitcoin protocol therefore has to solve a difficult problem: on the one hand, one should not be able to track individual users' activity, while, on the other hand, all transactions must be public. This problem is partially solved by the possibility for users to use as many addresses as they want. As a consequence, a same user can use a different address each time he receives a payment, and observers cannot, in theory, associate these different addresses to the single user that control them.

In practice, however, previous works have shown that it is possible, using some simple heuristics, to re-identify several addresses belonging to the same user.

In this article, we show how community detection, a well-known network analysis technique, can be used to improve on these simple heuristics to better re-identify users.

In section 2, we introduce key elements of the Bitcoin transaction system, necessary to understand how re-identification is made possible. In section 3, we present existing heuristics for user re-identification, while section 4 introduces our method based on community detection. In section 5, we propose a framework to empirically assess the quality of re-identification heuristics, and compare the proposed method to existing ones.
 
\section{Bitcoin transaction definition}
To explain how re-identification techniques are made possible, it is first necessary to explain some key elements on how the Bitcoin cryptocurrency works. In this article, we will not describe all details of the \textit{blockchain} and Bitcoin protocol, but only relevant information for the task of user re-identification. For more details on the protocol, one can refer to \cite{nakamoto2008bitcoin,garay2015bitcoin,kosba2016hawk}.

\bigskip

Bitcoin users can exchange amounts through \textit{transactions}, recorded in a public ledger, the \textit{blockchain}. These transactions are characterized by:
\begin{itemize}
	\item $m$ outputs, each associated with: \begin{itemize}
		\item A bitcoin amount
		\item A public key (address)
	\end{itemize}
	\item $n$ inputs, each corresponding to the output of a previous transaction (thus indirectly to its bitcoin amount and public key)
	\item A time stamp, which is an approximate value (due to the distributed nature of the blockchain, the timestamp is not exact, and in some cases, a transaction can take as input the output of a transaction with a later timestamp.
\end{itemize}

The sum of the inputs of a transaction is equal to the sum of the outputs plus an optional transaction fee.

Each address, which is in fact a \textit{public key}, is associated with a \textit{private key}. Both of them have been generated by the same user, and, while the public key is publicly shared, the pivate key is kept secret. To spend the bitcoins received at an address, its owner must "unlock" it by using its private key, as defined by the encryption protocol.

At this point, it is necessary to point out some important consequences of this mechanism:
\begin{itemize}
	\item The balance of a user $u$ at time $t$ is the sum of all transaction outputs that can be unlocked by a private key owned by $u$, and that have not yet been unlocked (i.e. spent) at time $t$.
	\item If two transactions $T_0$ and $T_1$ have the same address $d$ among their outputs, with respective associated amounts $a_1$ and $a_2$, these amounts are not merged or associated in any manner, i.e. there is no entity corresponding to the address $d$ that controls, or contains, $a_2+a_1$ bitcoins. Instead, these two outputs are uniquely identified respectively as output $m_0$ of transaction $T_0$ and output $m_1$ of transaction $T_1$, and each of these outputs can independently be used as input of a new transaction. Their only common trait is that they can be unlocked using the same private key, therefore by the same user.
	\item There is no partial spending: if a user who controls an output $o_1$ corresponding to $x$ bitcoins use it as input of a transaction to send $y$ bitcoins to another user, with $y<x$, then the user must send back $x-y$ bitcoins to an address that he/she controls.
	\item Users need to create multiple-input transactions each time they need to pay an amount superior to the value of each individual output they control. 
\end{itemize}

\section{Existing re-identification heuristics}
Several works have tackled the problem of users re-identification. In this section, we will describe heuristics based on transactions present in the public ledger. Some other methods exist, using IP-addresses collected in real time while transactions occur \cite{biryukov2014deanonymisation} or global activity patterns for very active users \cite{monaco2015identifying}. One can refer to \cite{lischke2016analyzing} for a recent survey on such methods. 

Methods based on the public ledger are the most common ones. They have the advantage of being based on data easily accessible to anyone.

\subsection{Heuristic based on transaction inputs only}
A first heuristic for re-identification has been introduced in \cite{reid2013analysis}: because all private keys corresponding to addresses must be used conjointly to sign a transaction, it can be safely assumed that all addresses in the input of a transaction belong to the same user. The introduction of the \textit{multisig} protocol in 2013 does not dramatically change this observation: it allows an output to be controlled by several users (all users must use their private key to sign), but does not allow to use conjointly outputs that were initially detained privately by different users. Furthermore, these conjointly detained addresses are identified differently from normal ones, and are scarcely used (as of 2017).

This heuristic can be defined as:
\begin{heur}
All addresses used as input of the same transaction belong to the same controlling entity, called a User.	
\label{H1}
\end{heur}

A simple network algorithm can be used to discover these users efficiently: 
\begin{enumerate}
	\item Create a network with nodes representing addresses
	\item For each set of $n$ addresses used conjointly as input of a transaction, add $n-1$ edges to create a path between them
	\item Discover connected components of this network.
\end{enumerate}
Connected components of this network correspond to sets of addresses belonging to a same user.

Despite the simplicity of this heuristic, it gives good results in practice, because users often have to use conjointly different addresses in order to perform transactions. Some users also tend to reuse the same addresses multiple times, either because they are not aware of the tracking possibility it allows, or because it is more convenient for them. For instance, the organization Wikileaks has an official address to which anyone can donate, therefore this address is used every time someone makes a donation to this organization. It is therefore possible to track how much money this organization receives through donations to this public address.

There is debate on the efficacy of this heuristic. Interested readers can refer to \cite{harrigan2016unreasonable} for a discussion on why this method gives good results in most cases.

%
%

\subsection{Heuristics based on inputs and outputs of transactions}
\textbf{H\ref{H1}} can only associate addresses because they simultaneously appear in the \textit{input} of a transaction. Another property of bitcoin is that, because users cannot do \textit{partial spending}, they have to send back to themselves the difference between the sum of inputs and the amount to pay. To improve the user discovery process, one needs to associate input addresses with one --or several-- output addresses that belong to the same user, called \textit{change address}, or \textit{shadow address}. Several authors have proposed heuristics to discover which output address might be a change address.

Existing heuristics are searching for so-called \textit{one-time change address}. Indeed, it is a common practice to receive the change on a newly generated address, that is never to be reused. 

Androulaki et al. \cite{androulaki2013evaluating} proposed the following heuristic to discover \textit{one-time change addresses}: 
\begin{heur}
If there are exactly two output-addresses $a1$ and $a2$, that one of them ($a1$) appears for the first time and that the other ($a2$) has appeared before, then $a1$ is considered to be the change address.
\end{heur}
This heuristic has also been used in \cite{spagnuolo2014bitiodine}.

Meiklejohn et al. \cite{meiklejohn2013fistful} propose a variation of this heuristic, not limited to transactions with exactly 2 outputs:
\begin{heur}
\label{Hmeiklejohn}
An address is considered a one-time change address if it satisfies the following properties:
\begin{itemize}
	\item The transaction is not a coin generation
	\item The address is not among the input addresses (address reuse)
	\item It is the only output address appearing for the first time
\end{itemize}	
\end{heur}
In their article, by using known addresses of several users as a reference, the authors are able to identify incorrectly grouped addresses, and propose a list of ad hoc restrictions to consider an address as a one-time change address, that involves checking if the address is not reused later, or reused only by some special services such as gambling games, that send their payout to the change address of the payment they received. The authors point out that these restrictions are only based on empirical observations, and cannot be generalized for future transactions. 

A simple adaptation of the network algorithm described for \textbf{H1} can be used for \textbf{H2} and \textbf{H3}: a network is created as for \textbf{H1}, and an edge is added between each identified one-time change address and one of the corresponding input addresses. Connected components of this graph correspond to set of addresses controlled by the same user. 

We can observe that there is a hierarchical relation between \textbf{H1} and \textbf{H2, H3}: each user of \textbf{H2, H3} corresponds exactly to one or several users in \textbf{H1}.

A more in-depth description of these approaches and their applications can be found in \cite{lischke2016analyzing}.

\section{Proposed Heuristic for user re-identification}
In this section, we propose to use community detection, a popular complex network analysis tool, to discover addresses belonging to a same user. The method is built on top of \textbf{H1} method, using its result as input.

The main idea is to create an \textit{identity hint network}, in which edges between nodes represent hints that the corresponding set of addresses might belong to a same user. Because community detection searches for dense subgraphs, sets of nodes with multiple hints of belonging to a same user will be grouped together by the algorithm. Thus this solution exploits weak signals to re-identify users.

We propose to create the \textit{identity hint network} using the following process: 
\begin{heur} -

\begin{enumerate}
	\item A first level of aggregation is created by applying \textbf{H1}. Sets of addresses belonging to a same user are used as nodes of the hint network
	\item For each transaction in the dataset, considering users found by \textbf{H1} instead of individual addresses, an edge is added (if not already present) between the (necessarily unique) sender and each recipient if:
	\begin{itemize}
		\item there are less than 10 users in the ouput of the transaction (recipients)
		\item all recipients are different from the sender, i.e there is no already known change address
	\end{itemize}
\end{enumerate}

On this network, a community detection algorithm is applied. Communities correspond to unique users.
\end{heur}

In our experiments, we used a popular hierarchical community detection algorithm, the Louvain algorithm \cite{blondel2008fast}, that yields several solutions at different levels of aggregation.

For a given level, we consider that a community corresponds to a user, and that this user controls all addresses associated with nodes in the cluster. As with \textbf{H2} and \textbf{H3}, users found by \textbf{H4} can be associated with one or more users of \textbf{H1}, i.e. there is a hierarchical relation between them.

\section{Evaluation of user discovery heuristics}
To the best of our knowledge, the quality of user re-identification heuristics has never been assessed empirically. In this section, we propose a method to do so, using tools from \textit{community detection} and \textit{clustering}.

Let's assume that we have at our disposal a ground truth defined as a set of $n$ labeled addresses $a_{GT}=\{a_1,a_2,\dots,a_n\}$, with $a^L$ the label of address $a$, corresponding to the user that controls it.

By using a heuristic on a transaction dataset, we discover sets of addresses identified as belonging to a same user. We assign a same label to all addresses of the same set, thus obtaining a set of $m$ labeled addresses, $a_{H}$. By filtering out all addresses of $a_{H}$ not present in $a_{GT}$, we obtain a set of addresses with two sets of labelling, the ground truth and a computed one. These labelling can be considered as clusters or network communities, and their similarity can be computed using, for instance, the NMI, chance adjusted NMI (aNMI) or Precision, Recall and $F_1$-score, which all give a different view on the tested clustering.

\subsection{NMI and aNMI}
The Normalized Mutual Information is frequently used to assess the correspondence between clustering or community partitions. It comes from information theory, and corresponds to the mutual information between the two compared partitions normalized between 0 --no similarity-- and 1 --identical partitions. Several variants of the normalization can be used; in this article, we use the following definition:

\[
NMI = \frac{I(U,V)}{\sqrt{H(U)H(V)}}
\]
With $I(U,V)$ the mutual information between partitions $U$ and $V$, and $H(U)$ the entropy (or quantity of information) of partition $U$. For more details on NMI, please refer to \cite{vinh2010information,lancichinetti2009detecting}.

Despite being often used in community detection evaluation, NMI suffers from known limitations, in particular that random partitions have expected values higher than 0, that can even be relatively high in some cases. A commonly used variant is the \textit{adjusted for chance} NMI, called aNMI, which conserves the normalization but correct for the effect of chance. Please refer to \cite{vinh2010information} for more information on aNMI.

\subsection{Precision, recall and $F_1$ score}
Precision and Recall are typically used to assess the quality of classification tasks. They can also be used to evaluate the quality of clustering: Precision corresponds to how often elements are correctly grouped together, while Recall evaluate how often elements belonging to a same cluster are labeled as such in the tested partition.

More formally, for each pair of addresses, we check if they are \textbf{TP},\textbf{FP} or \textbf{FN}: 
\begin{itemize}
\item \textbf{TP} (True Positive): two addresses with the same label are in the same cluster
\item \textbf{FP} (False Positive): two addresses with different labels are in a same cluster
\item \textbf{FN} (False Negative): two addresses with the same label are in different clusters
\end{itemize}
Then Precision and Recall are defined as usual: 
\[
Precision=\frac{TP}{TP+FP}, Recall=\frac{TP}{TP+FN}
\]
The $F_1$ score is the harmonic mean of Precision and Recall, and summarize the quality of a clustering in a single value.

\subsection{Test dataset}
In order to compare heuristics empirically, we need a labeled dataset of users and their known addresses. In the article \cite{meiklejohn2013fistful}, the authors have collected such a dataset by engaging in interactions with many bitcoin entities. In total, they engaged in 344 transactions with a variety of services, classified in 6 categories: Mining, Wallets, Exchanges, Vendors, Gambling and Miscellaneous. Please refer to the original article for more information on how the data was collected. After removing some particular cases, such as addresses of uncertain owners, or addresses controlled by the authors of the article, we were left with 776 addresses belonging to 90 different users. The user with the most known addresses, wallet service \textit{instawallet}, controls 151 of them, while we know only one address for 33 users.

This dataset, due to its manual collection, is of relatively limited size, but still allows us to test our method and observe broad differences between heuristics. In future works, it will be necessary to collect larger labeled datasets for robuster results.

For the bitcoin transaction dataset, we limited ourselves to the one used by the authors of the ground truth dataset \cite{meiklejohn2013fistful}. It is composed of 16,086,073 transactions and 12,056,684 distinct addresses.

\subsection{Results}
We computed NMI, aNMI, Precision, Recall and $F_1$-score for \textbf{H1}, \textbf{H3} and all hierarchical levels of our proposed method \textbf{H4}. Results can be found in Tab. \ref{tab:evalUsers}

\begin{table}
  \centering

  \begin{tabular}{c|ccccc}

   Heur. & Precision & Recall & $F_1$ & NMI & aNMI\\
    \hline
    H1 & \textbf{0.98} & \textit{0.77} & \textbf{0.86} & \textbf{0.89} & 0.65\\
    H3 & \textit{0.09} & 0.83 & \textit{0.16} & \textit{0.47} & \textit{0.15}\\
    H4-l1 & 0.75 & 0.79 & 0.77 & 0.86 & 0.66\\
    H4-l2 & 0.50 & 0.87 & 0.63 & 0.81 & \textbf{0.67}\\
    H4-l3 & 0.27 & 0.90 & 0.42 & 0.70 & 0.47\\
    H4-l4 & 0.25 & \textbf{0.91} & 0.39 & 0.67 & 0.43\\

  \end{tabular}

  \caption{Empirical comparison of User re-identification algorithms. \textbf{H4-l1 to H4-l4} correspond to the increasing hierarchical levels of the community detection based method (\textbf{H4}). Highest values in bold, lowest ones in italic.}
  \label{tab:evalUsers}

\end{table}

As expected, Precision score for the \textbf{H1} Heuristic is very high, which means that when two addresses are grouped together, they nearly always belong to the same user. By manually investigating the error, we found that addresses mistakingly grouped together belonged to two companies, one being in fact a branch of the other. More surprisingly, the Recall score is already relatively high for this heuristic. 

By adopting \textbf{H3}, which creates larger clusters, the Recall score rises. However, this gain is made at the cost of an important decrease in Precision, thus resulting in lower overall scores ($F_1$, NMI and aNMI)

We can observe that, as expected, our method reach higher Recall than the \textbf{H1} method, at the cost of a lower Precision. However, compared to \textbf{H3}, the loss in precision is lower. In particular, partitions of the highest level (\textbf{H4-l3},\textbf{H4-l4}) obtain higher scores than \textbf{H3} on all tested metrics (Pareto dominance). \textbf{H4-l1} and \textbf{H4-l2}, solutions of lower hierarchical levels, obtain highest aNMI score than the reference \textbf{H1} metric.

\subsection{Visual representation of re-identification}
To better understand the differences between different heuristics, we propose to use a visual representation.
We propose to use alluvial diagrams to compare re-identification algorithms results. In this visualization, each represented clustering is represented on a vertical axis; each address is represented by a horizontal, going from the cluster it belongs to in one clustering to the cluster it belongs to in the other clustering. A layout algorithm minimizes line crossings; horizontal lines with two common endpoints are grouped into a larger line, whose width corresponds to the sum of widths of addresses it represents.

To limit the visual complexity, we restrain the representation to one type of labeled users, which correspond to \textit{wallet services}. Results for other types of users leads to the same kinds of observations. 

\begin{figure*} []
\begin{center}
\includegraphics[width=0.9\textwidth]{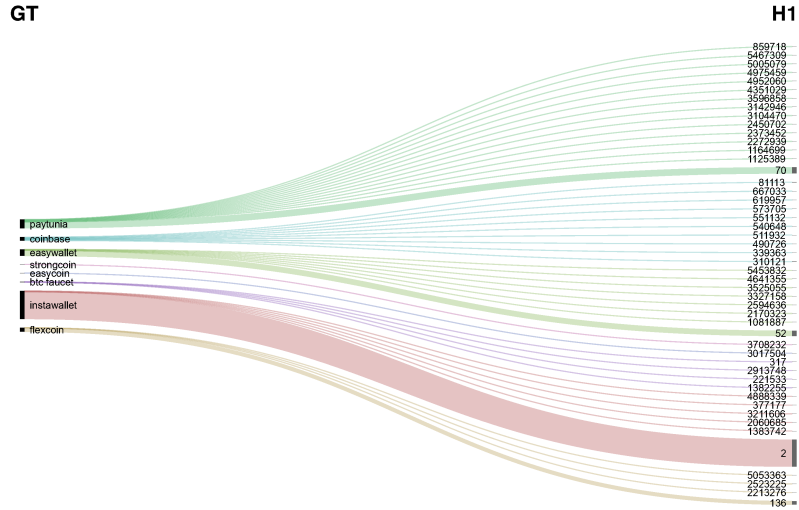}
\end{center}
\caption{Alluvial diagram representing the matching between clusters of addresses in the ground truth and \textbf{H1}. Each horizontal line corresponds to an address. A thick line corresponds to the aggregation of a set of addresses with common left and right belongings. We can observe that some entities (e.g \textit{instawallet}) are already quite well identified by this heuristic, for the given test set. Nevertheless, for most entities, their addresses are divided into several clusters in H1.}
\label{fig:H1}
\end{figure*}

Fig. \ref{fig:H1} represents the correspondence between clusters in the ground truth and H1. We can observe that users controlling large sets of addresses tend to be well discovered (large horizontal lines for i\textit{nstawallet, easywallet, paytunia, flexcoin}). We can also observe that there is no cluster in \textbf{H1} with addresses belonging to different users in the ground truth.

\begin{figure*} []
\begin{center}
\includegraphics[width=0.9\textwidth]{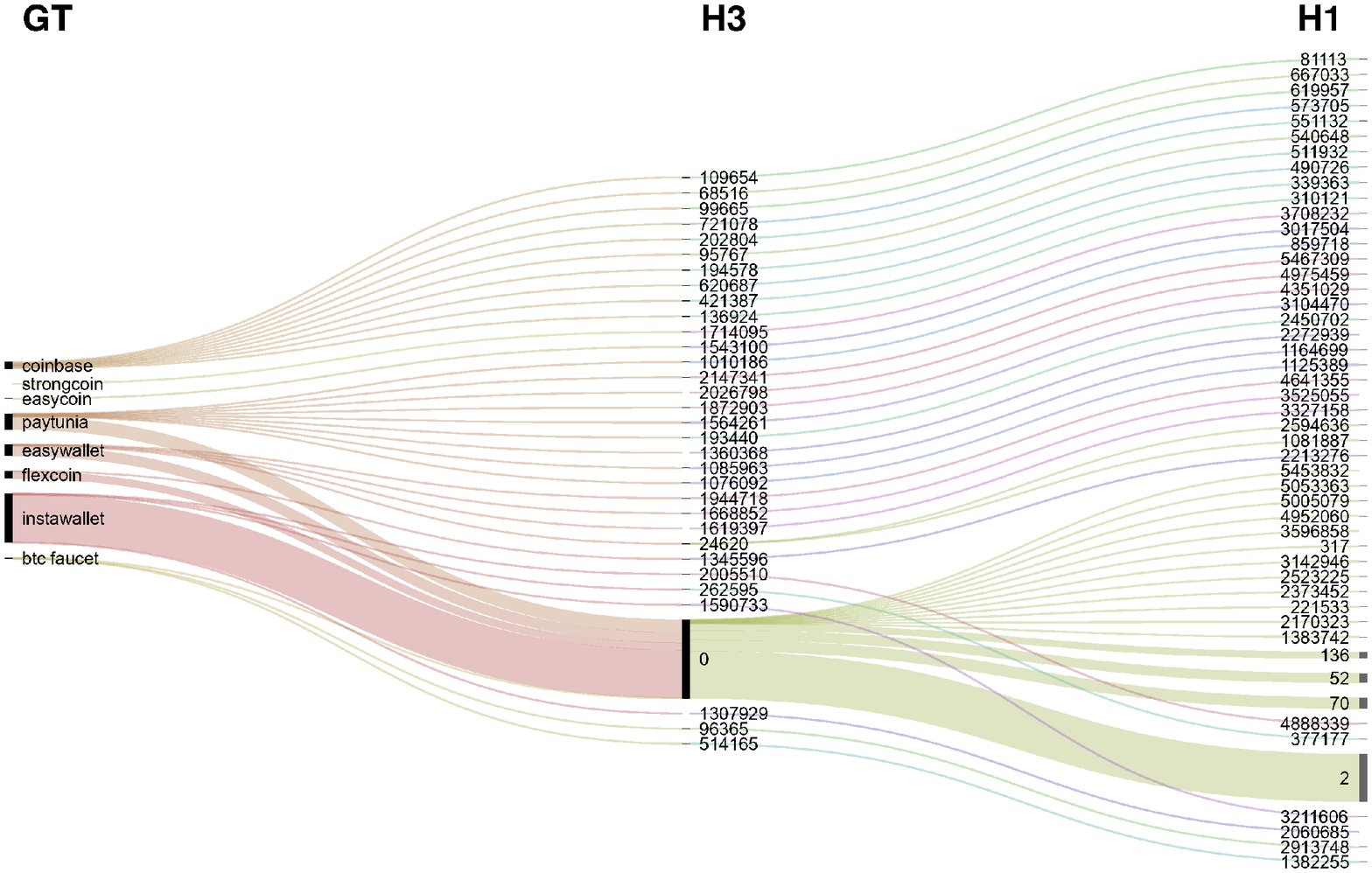}
\end{center}
\caption{Alluvial diagram representing the matching between clusters of addresses in the ground truth, H3 and H1. We can observe the apparition of a large cluster (named 0) that merge together many addresses corresponding in fact to different entities (Paytunia, easywallet, flexcoin, instawallet, etc.).}
\label{fig:H3}
\end{figure*}

Fig. \ref{fig:H3} represents the correspondence between clusters of addresses between the ground truth, \textbf{H1} and \textbf{H3}. We can observe in \textbf{H3} the presence of a large cluster (0) gathering addresses from different users in the ground truth. We can observe one occurrence (cluster 24620 in \textbf{H3}) of a cluster in \textbf{H3} rightfully grouping together distinct clusters from \textbf{H1}. 

\begin{figure*} []
\begin{center}
\includegraphics[width=0.9\textwidth]{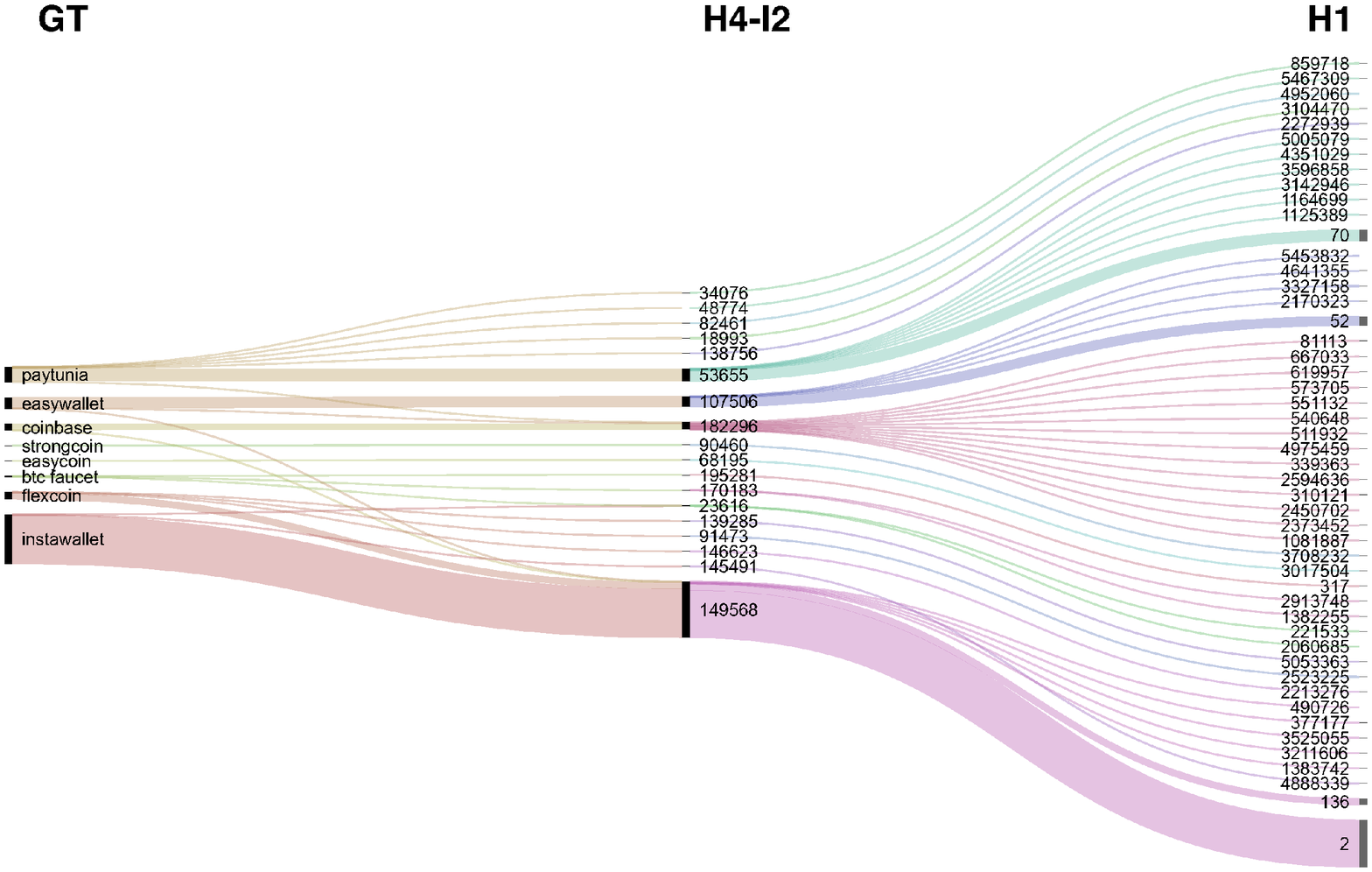}
\end{center}
\caption{Alluvial diagram representing the matching between clusters of addresses in the ground truth, H3 and H1. We can observe the apparition of a large cluster (named 0) that merge together many addresses corresponding in fact to different entities (Paytunia, easywallet, flexcoin, instawallet, etc.).}
\label{fig:H4}
\end{figure*}

Fig. \ref{fig:H4} represents the matching between clusters of addresses in the ground truth, \textbf{H1} and \textbf{H4-l2}. We can observe that some entities in the ground truth (\textit{flexcoin}, \textit{instawallet}, parts of \textit{coinbase} and \textit{easywallet}), are erronously merged in \textbf{H4-l2}, but less than with \textbf{H3}, as shown in Fig. \ref{fig:H3}. On the contrary, we observe several cases where different clusters in \textbf{H1} are correctly merged by \textbf{H4-l2} into clusters very close to those of the ground truth (for \textit{paytunia}, \textit{easywallet} and \textit{coinbase}, in particular)

We can therefore conclude that the proposed method offer a better solution than the \textbf{H3} method, and even provide a relevant alternative to the widespread \textbf{H1} solution. 

\subsection{Advantages of the proposed method}
On top of yielding a relevant user discovery solution, the community detection approach has several advantages:
\begin{itemize}
	\item Because it allows to extract several hierarchical levels, one can choose, depending on the application, to increase Recall at the cost of Precision, or said differently, to discover more addresses for known users, at the risk of mistakingly merging different users.
	\item Because the solution is not based on a set of rules, but on \textit{weak signals}, it is more difficult for users to develop countermeasures to escape re-identification of their addresses
	\item The proposed solution is a generic framework that can be adapted in countless ways to be improved: one could create a different \textit{hint network}, using weighted edges based on rules decided by experts, or based on machine learning. One could also use other community detection algorithms. Finally, one could also take into account the dynamic of transactions, through temporal networks \cite{holme2012temporal} and dynamic community detection \cite{cazabet2014dynamic}.
\end{itemize}

\section{Conclusion}
In this article, we have shown how community detection could be used to improve re-identification of users in bitcoins. 

We think that this article has two main contributions: on the one hand, it provides a framework for empirical testing and comparison of re-identification algorithms. The dataset used as ground truth is limited in size and representativeness, and further work is needed to produce more ambitious ground truth datasets in order to validate algorithms on a large scale.

On the other hand, we have shown how using complex networks analysis tools such as community detection could improve existing techniques, based on set of rules. In particular, contrary to set of rules techniques, they cannot be easily bypassed by users when publicly released.

\begin{acknowledgement}
We want to thank the authors of \cite{meiklejohn2013fistful}, and in particular Sarah Meiklejohn, for kindly sharing with us their dataset, result of their remarkable work.

This work is funded in part by the European Commission H2020 FETPROACT 2016-2017 program under grant 732942 (ODYCCEUS), by the ANR (French National Agency of Research) under grants ANR-15-CE38-0001 (AlgoDiv) and ANR-13-CORD-0017-01 (CODDDE), by the French program “PIA - Usages, services et contenus innovants" under grant O18062-44430 (REQUEST), and by the Ile-de-France program FUI21 under grant 16010629 (iTRAC).\end{acknowledgement}

\bibliographystyle{spmpsci}
\bibliography{refBitcoin}
\end{document}